\documentclass[twocolumn]{aastex701}
\usepackage{graphicx}

\usepackage{ulem}
\usepackage{color,soul}

\begin{document}
\title{Watching a Comet Turn On: High Spectral Resolving Power Observations
of Comet C/2017 K2 (PanSTARRS) }
\correspondingauthor{Anita Cochran}
\author[0000-0003-4828-7787]{Anita L. Cochran} 
\email{anita@astro.as.utexas.edu}
\affil{McDonald Observatory, University of Texas at Austin}
\author[0000-0002-0622-2400]{Adam J. McKay}
\email{mckayaj@appstate.edu}
\affil{Appalachian State University}
\author[0000-0001-9784-6886]{Youssef Moulane}
\email{youssef.moulane@um6p.ma}
\affil{School of Applied and Engineering Physics, Mohammed VI Polytechnic University}

\begin{abstract}
We report high spectral resolving power optical observations of comet C/2017\,K2 (PanSTARRS) as it approached the Sun.  This comet was discovered when it was 16\,{\sc au} from the Sun. At discovery, the comet had a large and relatively bright coma. However, the spectrum at discovery showed only signatures of dust. We used the coud{\'e} spectrograph on the McDonald Observatory 2.7\,m telescope to obtain spectra, starting when the comet was at
a heliocentric distance of 5.06\,{\sc au} and following it until 2.47\,{\sc au}, to determine what spectral features would appear at each heliocentric distance.  The first heliocentric distance for which we detected any emission from the gas was 
3.39\,{\sc au}, when we first detected CN.  As the comet continued inward towards the Sun, various other species were detected. We discuss the implications of the early turn-on of CN and of species first appearing at different heliocentric distances in the context of control of the activity by water.

\end{abstract}

\keywords{\uat{Comae}{271} --- \uat{Comets}{230} --- \uat{Long period comets}{933} --- \uat{High resolution spectroscopy}{2096}}

\section{Introduction}
Comets are leftover building blocks from the formation of the Solar
System.  They have a gas-to-dust mass ratio of approximately
1, with computed ranges from $0.5\rightarrow>10$ \citep{patzold2019,chetalssrv2020,Marschalletal2020}. As small bodies that spend most of their lives far from the
Sun, the ices stay frozen for most of their existence.  However, on
occasion, comets get perturbed into the inner Solar System.  As they
approach the Sun, they are heated, and the ices will sublime.  The small
size of these bodies means they will not have strong gravity fields, so the gas released from the sublimation will flow away from the nucleus.  This gas will carry some of the dust.

The amount of heating needed to cause sublimation depends on the type and quantity of the ice.  The ices in most cometary nuclei are about 80\% H$_{2}$O and 20\% other ices.  CO and CO$_2$ are next most important after water, while other species are mostly trace elements.  CO and CO$_2$ are more volatile
than H$_{2}$O and should sublime further from the Sun.  However, with
H$_{2}$O ice being so dominant, H$_{2}$O generally controls most of the ice
sublimation.  H$_{2}$O stays in its ice form further from the Sun than
3.0\,{\sc au}. Thus, most of the activity of a comet nucleus should occur
within $\sim$3.0\,{\sc au} of the Sun.

This general picture of activity turn on for a comet is difficult to
test since most comets are not bright at large heliocentric distances
and therefore are not discovered before the sublimation point of H$_{2}$O
ice.  
However, comet 2017~K2 (PanSTARRS) (hereafter K2) is a rare exception.
It was discovered on 21 May 2017 with the PanSTARRS
telescope. At discovery, it was at a heliocentric distance of
$\sim$16\,{\sc au} and it was already quite bright and extended. At such
a large distance from the Sun, H$_{2}$O sublimation should not have
started. So, what was causing the large coma seen at the time of discovery? We used high spectral resolving power optical observations to try to answer this question. 

Comet spectra have two components.  The gas that has sublimed from the
ices is fluoresced by sunlight and produces an emission line spectrum.
This spectrum can be very rich when the comet is near the Sun
\citep{coco02atlas, Biver2024CometBook}. The dust that is flowing out of the coma reflects the
light from the Sun, producing an absorption spectrum, similar to the Sun's
spectrum, but shifted from the telescope rest frame based on the geocentric
radial velocity of the comet.  Only the dust
spectrum was observed from the large, bright active coma when
the comet was first discovered far from the Sun.

However, CO was detected when the comet reached 6.7\,{\sc au} \citep{yangetal2021}, while CO or CO$_2$ were detected at even larger distances with NEOWISE (CO and CO$_2$ are in the same bandbass for NEOWISE and cannot be differentiated.) \citep{Miletal2024}.
Once gas starts to sublime, 
the emission line spectrum from the gasses is more indicative of what
controls the activity of the comet. We monitored the comet as it approached the Sun to determine which molecular species appeared first in the spectrum and at what heliocentric distances.

With emission line spectra, weak features can be seen more readily with
high spectral resolving power observations than with lower resolving power.
This is because for the same equivalent width of a feature, low resolving power instruments spread the light over a much wider bandpass, resulting in the
peak of a feature being less pronounced and hidden in the noise in the
continuum.  For high spectral resolving power, more of the equivalent width
of a feature will be above the level of the continuum. Thus, narrow molecular
line features will stick up above the continuum for high resolving power and be
lost at low resolving power.

\section{Observations and Reductions}

We obtained spectra of K2 pre-perihelion using the Tull coud\'{e} spectrograph \citep{TuMQSn95} on the  Harlan J. Smith 2.7\,m telescope of McDonald Observatory on 8 nights in 2021 and 2022.
In addition, we observed the comet post-perihelion in 2023, once at
McDonald with the Tull spectrograph and once at Keck with HIRESb
\citep{vogthiresspie94}. Perihelion for K2 was on 19 December 2022, with the comet at 1.8\,{\sc au}. However, the comet moved too far south and too close to the Sun for us to observe it from McDonald Observatory after mid-August 2022. Indeed, at perihelion the comet had a declination of -60$^\circ$, or 90$^\circ$ south of McDonald Observatory. Details of the observations are given in Table~\ref{log}.

\begin{table*}
\caption{Log of Observations 
\label{log}}
\smallskip
\hspace{-1cm}
\begin{tabular}{r@{\,}l@{\,}lcccccc}
\hline
\multicolumn{3}{c}{Date (UT)} & Heliocentric & Geocentric & Heliocentric & Geocentric & Num. & Total \\
& & & Distance (AU) & Distance (AU) & Radial Velocity & Radial Velocity & Spectra & Exposure \\
& & & & & (km/sec) & (km/sec) & & (sec) \\
\hline
18 & Oct & 2021 & 5.06 & 5.45 & -15.05 & 1.14 & 3 & 4,800 \\
15 & Jan & 2022 & 4.27 & 4.96 & -15.52 & -22.37 & 2 & 1,800 \\
22 & Apr & 2022 & 3.39 & 3.00 & -15.69 & -40.05 & 10 & 12,000 \\
09 & May & 2022 & 3.24 & 2.64 & -15.62 & -37.66 & 11 &12,440 \\
10 & Jun & 2022 & 2.95 & 2.05 & -15.34 & -24.67 & 14 & 16,800 \\
25 & Jun & 2022 & 2.82 & 1.89 & -15.12 & -14.56 & 11 & 13,200 \\
31 & Jul & 2022 & 2.51 & 1.86 & -14.19 & 9.86 & 3 & 3,600 \\
06 & Aug & 2022 & 2.47 & 1.90 & -13.98 & 12.50 & 8 & 9,600 \\
06 & Nov & 2023$^a$    & 4.11 & 3.56 & 15.59 & -7.42 & 4 & 3,600 \\
16 & Nov & 2023 & 4.20 & 3.52 & 15.55 & -4.10 & 3 & 1,800 \\
\hline
\multicolumn{7}{l}{$^a$ Observed with Keck HIRESb; all others McDonald Tull Spectrograph} \\
\end{tabular}
\end{table*}

The Tull spectrograph was used with a 1.2 arcsec wide by 8.2 tall slit centered on the optocenter for all McDonald observations,
yielding a resolving power ($R=\lambda/\Delta\lambda$) of 60,000. 
The HIRESb spectrograph was used with a 0.86 arcsec wide by 7 arcsec tall slit,
yielding a resolving power of R=48,000.

On all dates, we obtained a number of spectra of the comet and combined
them during the reduction.
As long as the detected signal was above the detector read noise, there is no
penalty (except for the short readout time) to using short exposure times.
However, with many short exposures, we can
combine subgroups of the data using a median combination to
remove ``cosmic rays'' (energetic particle events). 
Since we anticipate our detections to be emission
lines, this allows us to reject cosmic rays without rejecting real
features.  Then we can sum all the groups to create the most accurate
spectra possible. 

Data reduction followed standard processes for the spectrographs.  Bias
frames were used to create a master to remove the bias offset in the images. Summed spectra of incandescent lamps were used to define the spectral path across the detector of each order, the width of each order and to use as a flat field to remove the grating blaze. 
Any scattered light from the flat field was removed by fitting a surface
under the spectral image.  Scattered light was also removed from the comet and star spectra.  Then, the object spectra were corrected for the flat field. At this point, the spectra were extracted from 2D to 1D using the tracks defined either with the flat field (for fainter targets) or the actual object spectral image (brighter targets). ThAr lamp spectra were processed similarly. Using the ThAr spectra, we were able to identify the wavelengths of every point on the 1D spectra with an accuracy of 0.002\AA. 
At the end of the routine processing, we had spectra that consisted of about 63 spectral orders for the Tull spectrograph and 53 orders split over 3 detectors for HIRES. 
The McDonald Observatory data presented in this paper will be made available by the NASA
Planetary Data System under the data collection identifier
$urn:nasa:pds:gbo-mcdonald:comet\_echelle\_survey$ (Cochran, submitted). The Keck spectra can be found in the Keck Archive.

As noted above, the spectra consist of molecular emission features
superposed on a solar continuum and absorption spectrum resulting from 
solar light reflected by the coma dust. In order to search for weak gas features, we first needed to remove the dust spectrum.  We did this using a solar spectrum observed with the same spectrographs.
For the McDonald observations, the solar spectrum was obtained by imaging
daylight sky through a diffusing window into the spectrograph along the
same path from the slit to the detector that star light or comet light
follows.  Solar spectra were obtained each afternoon prior to the comet
observations.  For the HIRES observations, no such solar port exists.
Therefore, we observed the solar analogue star HD28099 = Hyades 64.

\begin{figure}
\includegraphics[width=0.45\textwidth]{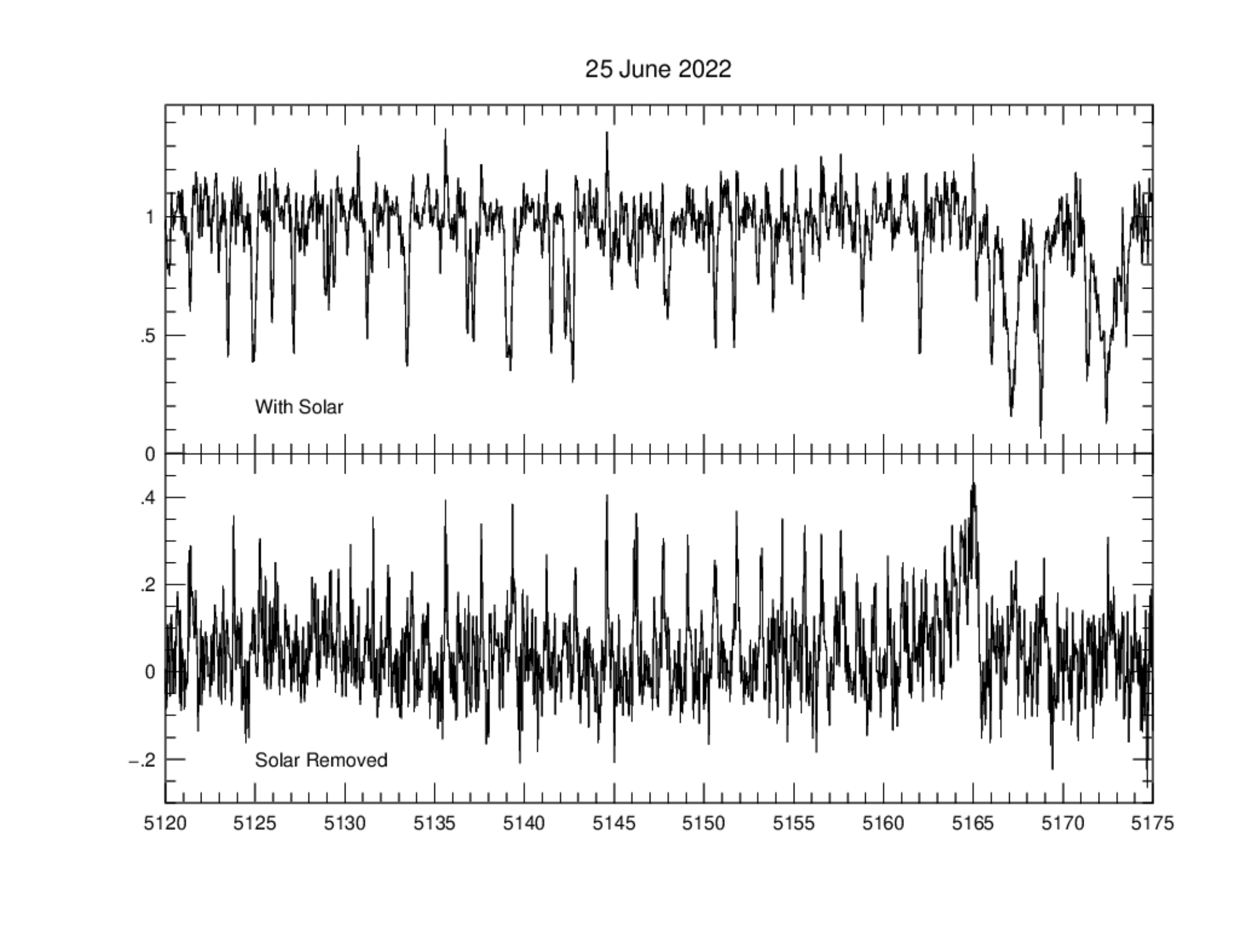}
\caption{An example of a spectrum from 25 Jun 2022 in the region of the
C$_{2}$ (0,0) bandhead is shown. The top panel is the spectrum with the
molecular emission spectrum plus the absorption and continuum from dust reflected sunlight. The bottom panel
shows that spectrum with the solar contribution removed.  In the lower panel,
many emission lines are quite obvious that cannot be seen in the upper panel,
including the C$_{2}$ (0,0) bandhead at $\sim$5165\AA.
\label{c2remove}}
\end{figure}

Searching for features required us to work on a single order at a time.
To remove the solar spectrum from the comet spectrum,
we first shifted the solar spectrum in wavelength and scaled it in counts until the comet and solar spectra were approximately matched at a continuum wavelength region near where there was a feature of interest. The RMS was then interactively minimized by shifting the solar spectrum in counts and wavelength.
At that point, we subtracted the scaled and shifted solar spectrum from the
comet spectrum. This left us with spectra of the potential molecular emission lines of the comet without dust continuum.  We then applied a Doppler correction to the comet molecular spectra to put them on the telescope rest frame.
Figure~\ref{c2remove} shows a spectrum of the (0,0) bandhead region of C$_{2}$ from 25 June 2022 before and after the solar spectrum has been removed. After removal of the solar spectrum, the
C$_{2}$ bandhead and many C$_{2}$ lines are apparent. It would have been
difficult to pinpoint the C$_{2}$ lines without the solar spectrum being removed.

\section{Changes in the Spectra as the Comet Approached and Receded from the Sun}
\begin{figure}
\centering
\includegraphics[width=0.46\textwidth]{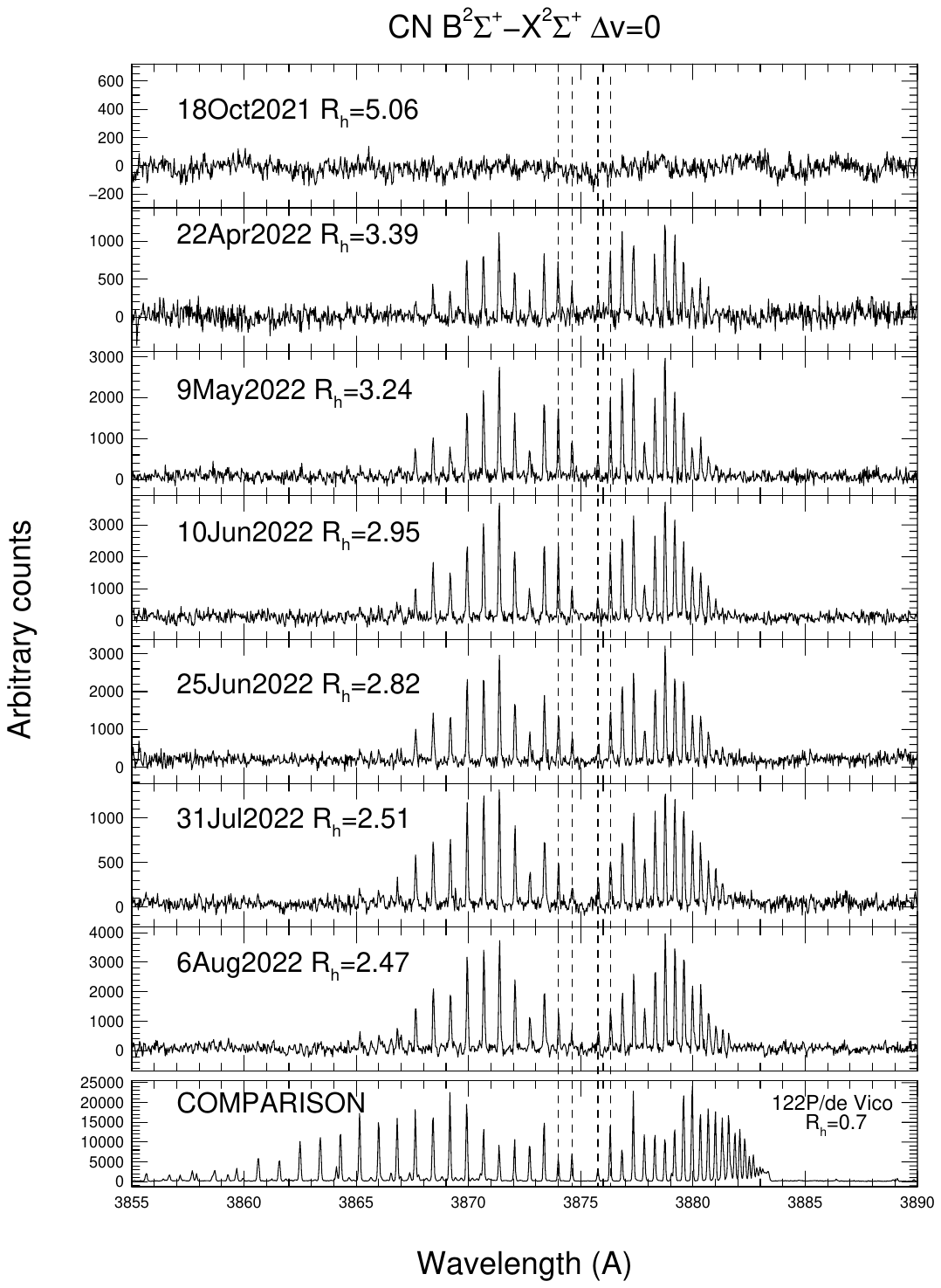}
\caption{The CN B$^2\Sigma^+$ -- X$^2\Sigma^+$  order of
the spectra for each pre-perihelion night except 15 Jan 2022 are
shown. There are clearly no CN lines in the spectrum from 18 Oct 2021 when the comet was at a heliocentric distance of 5.06\,{\sc au}.  By the time the comet is at 3.39\,{\sc au} on 22 Apr 2022, the CN band is well developed.
The dashed lines indicate the lowest j-levels of the P branch (right 2) and of the R branch (left 2).
\label{CNall}}
\end{figure}

\begin{figure}
\centering
\includegraphics[width=0.46\textwidth]{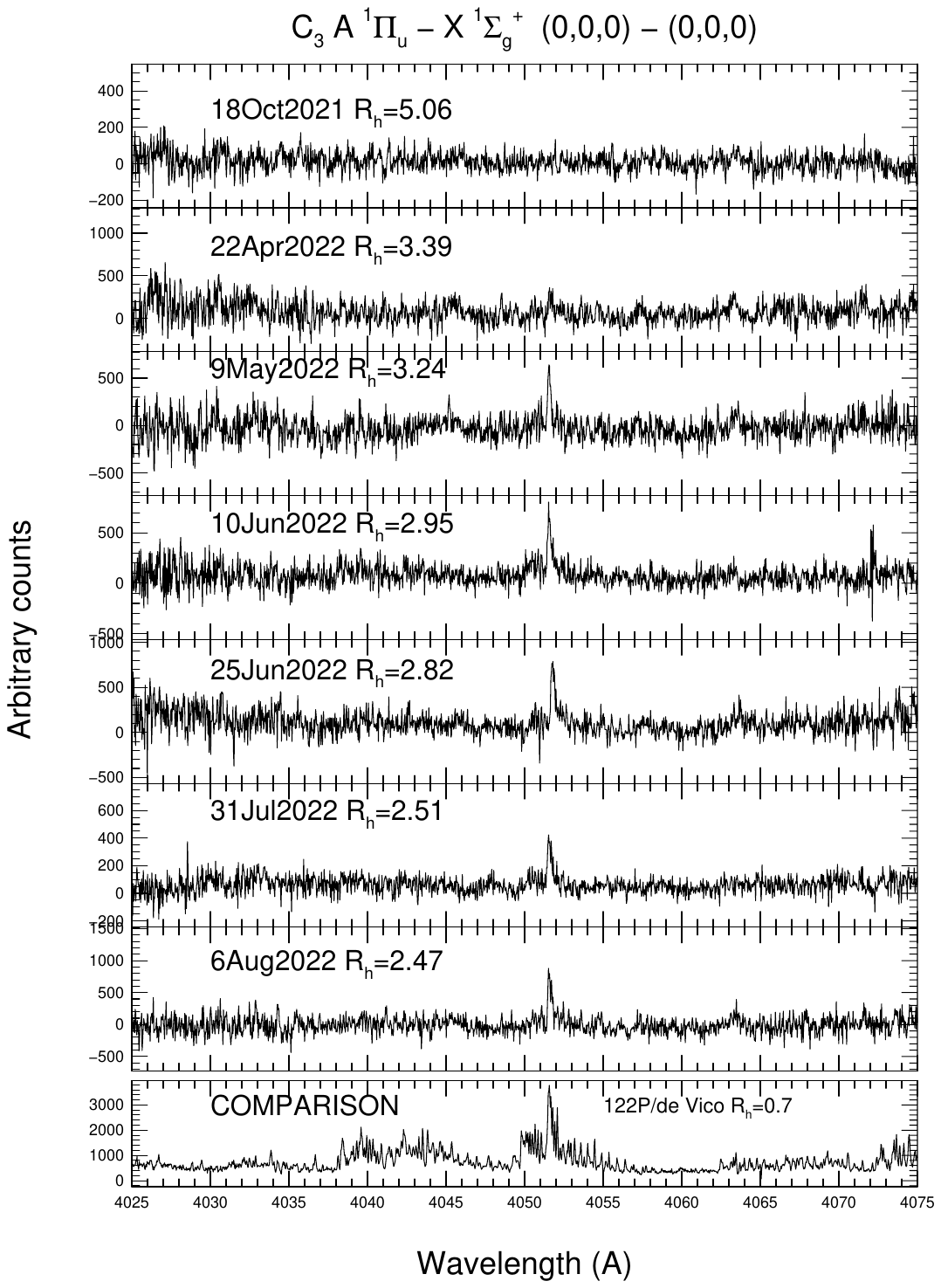}
\caption{The C$_{3}$ A$^1\Pi_u$ -- X$^1\Sigma_g^1$ (0,0,0) - (0,0,0) order of
the
spectra for each pre-perihelion night except 15 Jan 2022 are
shown. There are clearly no C$_{3}$ lines in the spectrum from 18 Oct 2021 
at a heliocentric distance of 5.06\,{\sc au} nor at R$_h$=3.39\,{\sc au} on 22
Apr 2022.   By 9 May 2022 at 3.24\,{\sc au}, the Q-branch bandhead at 4050\AA\ is clearly visible.
\label{C3all}}
\end{figure}

\begin{figure}
\centering
\includegraphics[width=0.47\textwidth]{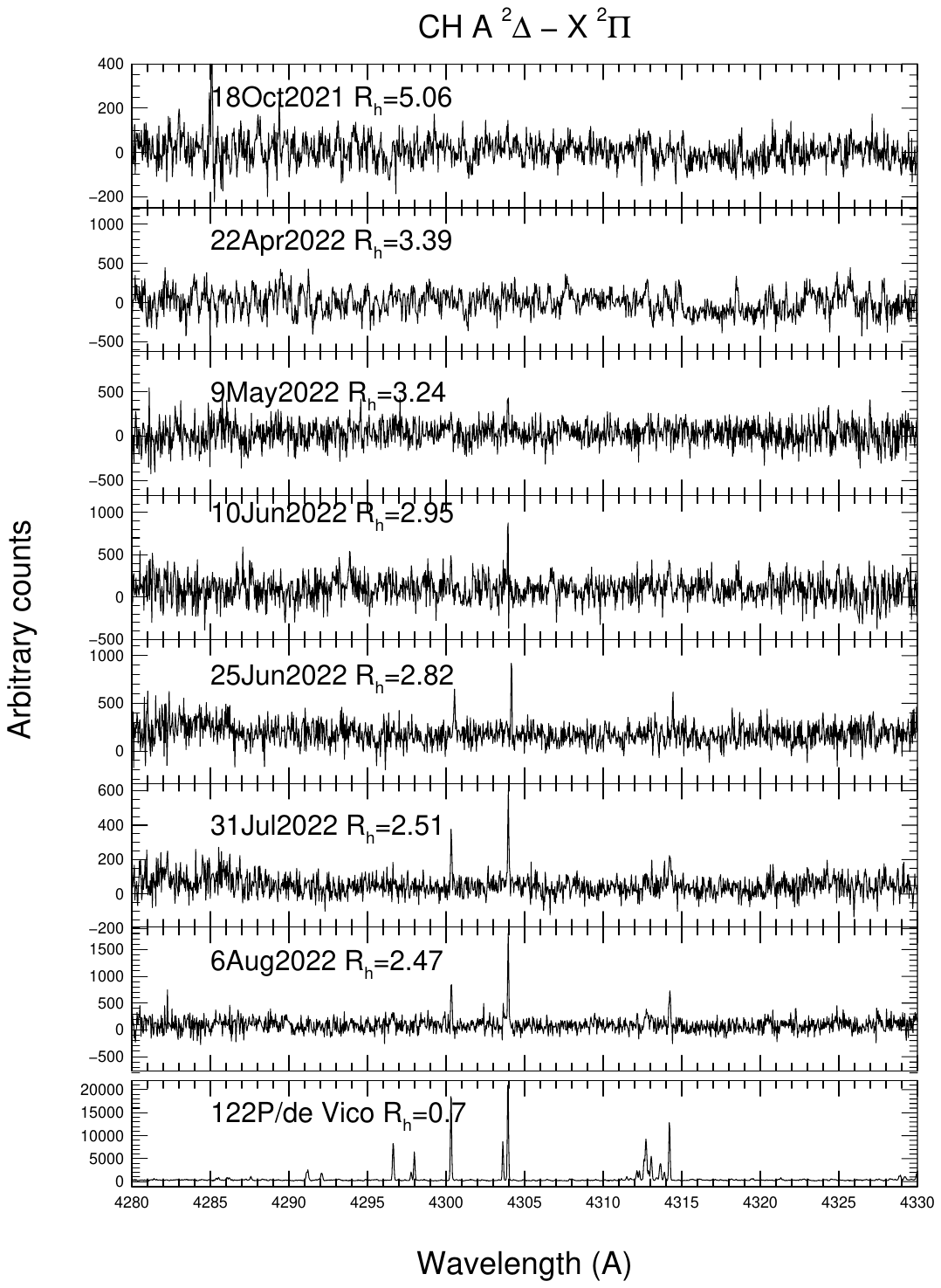}
\caption{The CH A$^2\Delta$ -- X$^2\Pi$ band does not
have many lines.  No CH lines are visible until 10 Jun 2022 at R$_h$=2.95\,{\sc au}.
The spectrum from 9 May 2022 is ambiguous for a line detection.
\label{CHall}}
\end{figure}

\begin{figure}
\centering
\includegraphics[width=0.47\textwidth]{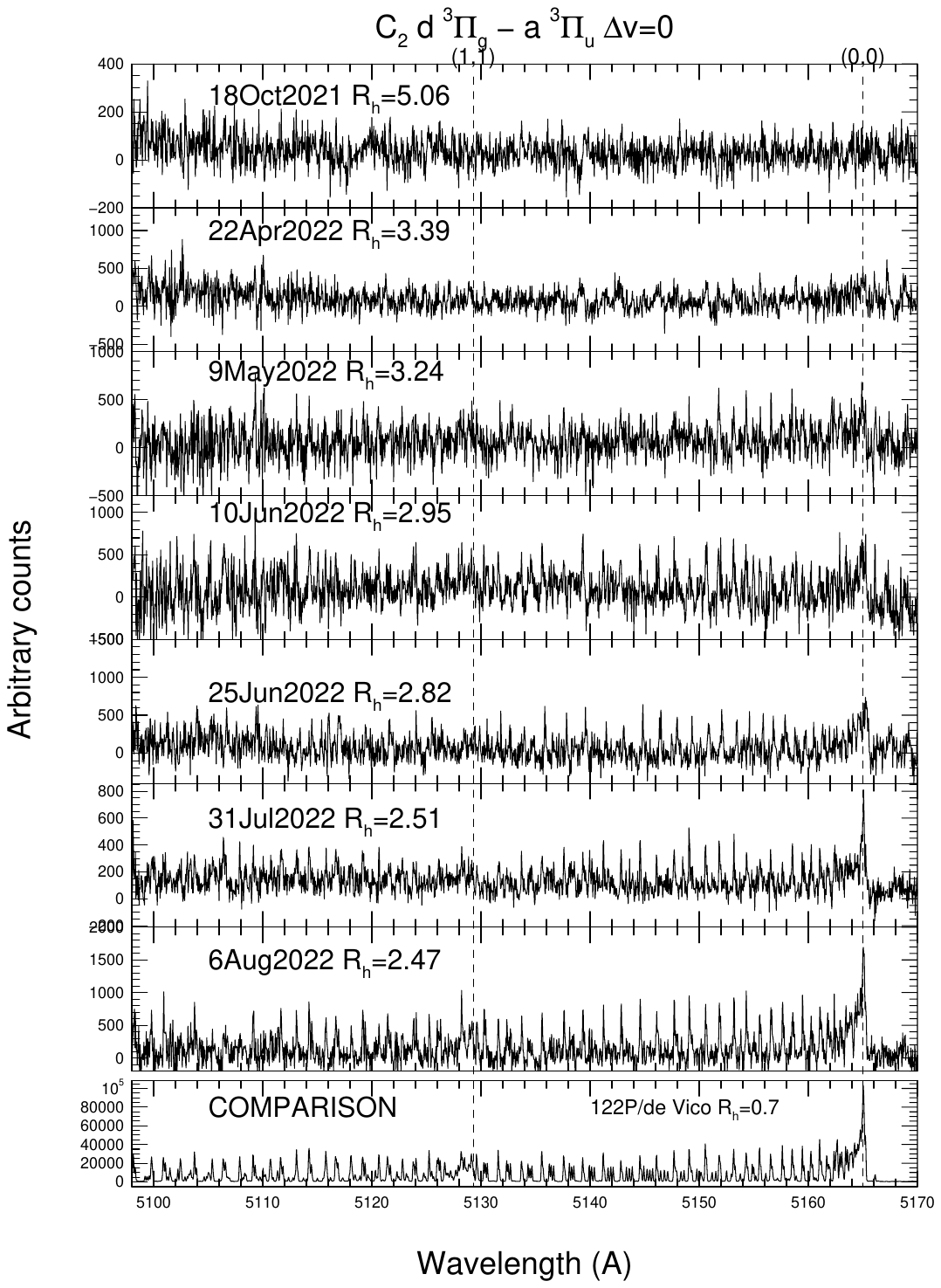}
\caption{The C$_{2}$  d$^3\Pi_g$ -- a$^3\Pi_u$ $\Delta v=0$ order of the
spectra for each pre-perihelion night except 15 Jan 2022 are
shown. There are clearly no C$_{2}$ lines in the spectrum from 18 Oct 2021 
at a heliocentric distance of 5.06\,{\sc au}.  The spectrum from 22 April 2022
probably shows no features, while the (0,0) bandhead is possibly visible on
9 May 2022.  The (0,0) bandhead is obviously present on 10 Jun 2022 while
the (1,1) bandhead begins to be believable on 31 Jul 2022. The two
bandheads are marked with dashed lines.  \label{allc2} }
\end{figure}

\begin{figure}
\centering
\includegraphics[width=0.475\textwidth]{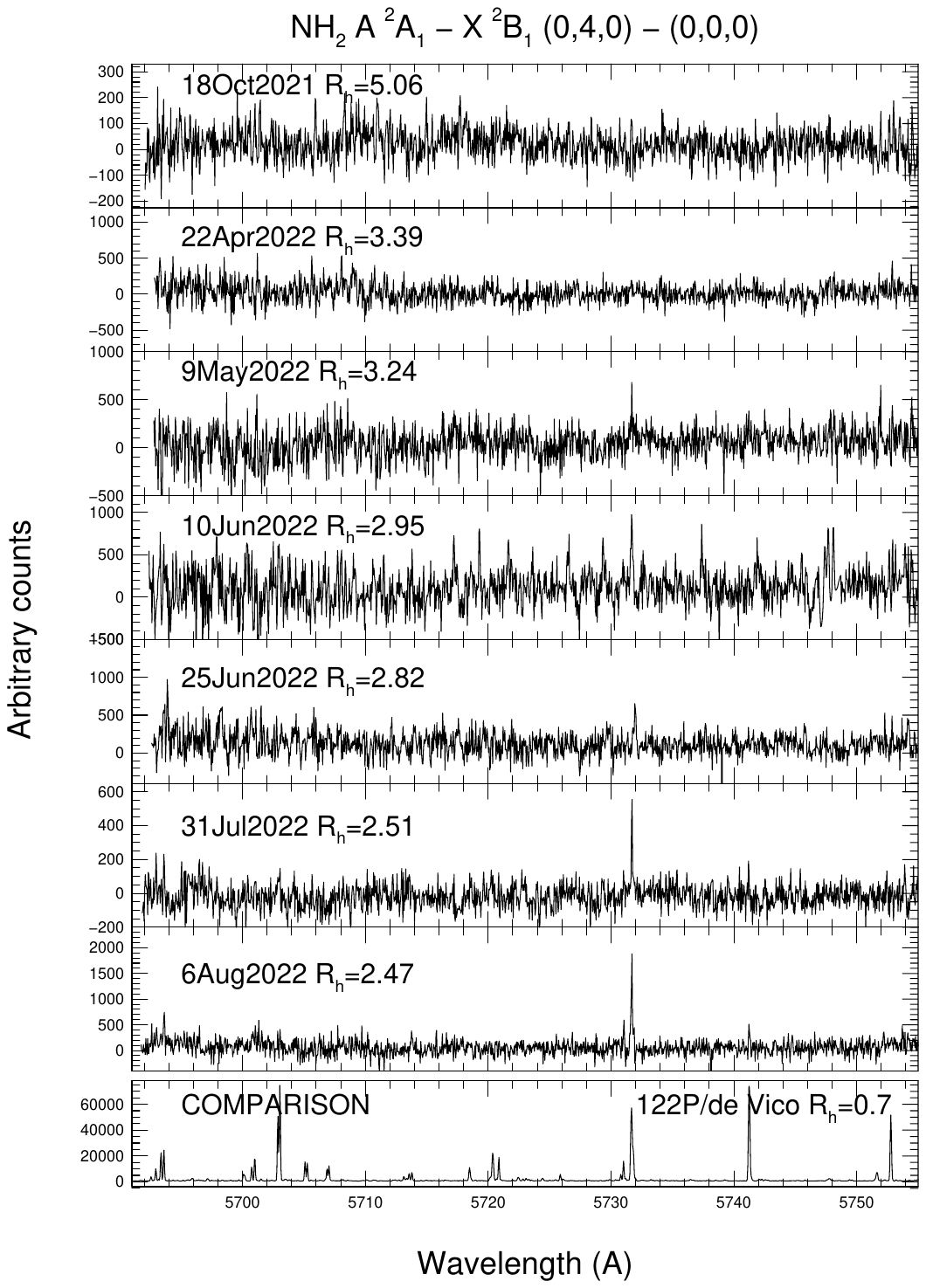}
\caption{The NH$_2$ A $^2$A$_1$ --X $^2$B$_1$ (0,4,0) -- (0,0,0) of NH$_{2}$ is
shown.  In general, the NH$_{2}$ lines do not form obvious band structures.
The line at around 5731.5\,\AA\ is the most likely line to detect in this bandpass. 
It is clearly there at R$_h$=2.51\,{\sc au} and may be detected as far out as 3.24\,{\sc au}.
\label{allnh2}}
\end{figure}

The procedure outlined in the previous section was applied to the
pre-perihelion spectra on each date (except 15 Jan 2022, see below) for
CN, C$_{3}$, CH,
C$_{2}$, and NH$_{2}$. Figures~\ref{CNall} -- \ref{allnh2} show the most
sensitive part of the spectrum for each of the molecules.  The spectra are
displayed from the earliest (largest heliocentric distance) to the spectrum
obtained closest to the Sun.  Each night's spectrum is in its own panel with date and
distance information included.  The y-axis is an arbitrary count that is used to
display any features in the full Y-range.
The scales differ
because of the different exposure times (and hence overall signal). All of
the spectra are displayed on the rest frame of the telescope. 

Below the
comet spectra for each molecule is an additional panel (separated slightly
from the K2 spectra) of comet 122P/de~Vico obtained with the Tull
spectrograph in 1995 \citep{coco02atlas}.  These were obtained when de~Vico was much closer to the
Sun than any of the K2 spectra.  This was also a very bright comet, so the
signal/noise is much higher than for any K2 spectra.  The de~Vico spectra {\it
do not} have the solar spectrum removed since de~Vico had very little dust
and therefore very little solar spectrum. Removal of the solar spectrum
would potentially add noise to the de~Vico spectra shown.  Comparison of
the K2 spectra and the de~Vico spectrum for each molecule helps confirm
detections in the K2 spectra.  Features seen in the K2 spectra, but not the
de~Vico spectra, are probably not real.

We could not use the data from 15 Jan 2022 for most of the molecules
because the underlying absorption/continuum spectrum on that night was
peculiar and could not be removed in the normal manner.  It turns out that,
coincidentally, the comet was at the same position on the sky as a galaxy, and the comet was more extended than that galaxy. 
Thus, we did not see the
interloper for our spectra with the guide camera, but the galaxy spectrum
was imprinted on top of the comet spectrum.  
Unfortunately, we did not realize that this interloper existed at the time and
did not get a separate spectrum of the galaxy.

The spectra of galaxies generally show a series of hydrogen line
absorptions from the Balmer sequence, with H$\alpha$ at 6563\,\AA, $\beta$ at
4861\,\AA, $\gamma$ at 4340\,\AA, and subsequent lines continuing into the
blue/UV at ever narrower intervals.  Because these are absorption lines,
galaxy spectra get weaker to the blue through the Balmer break at
3645\AA.  Thus, the galaxy's contribution to the underlying continuum
spectrum of the comet on 15 Jan 2022 decreased below $\sim$3900\AA\ 
to the point
that it contributed very little to the comet continuum.  For this reason,
though we could not remove the galaxy spectrum from most of the comet
spectra on 15 Jan 2022, we were able to try to remove the continuum below
3900\AA, or in the region we expect the CN band.

\begin{figure*}
\centering
\includegraphics[scale=0.65]{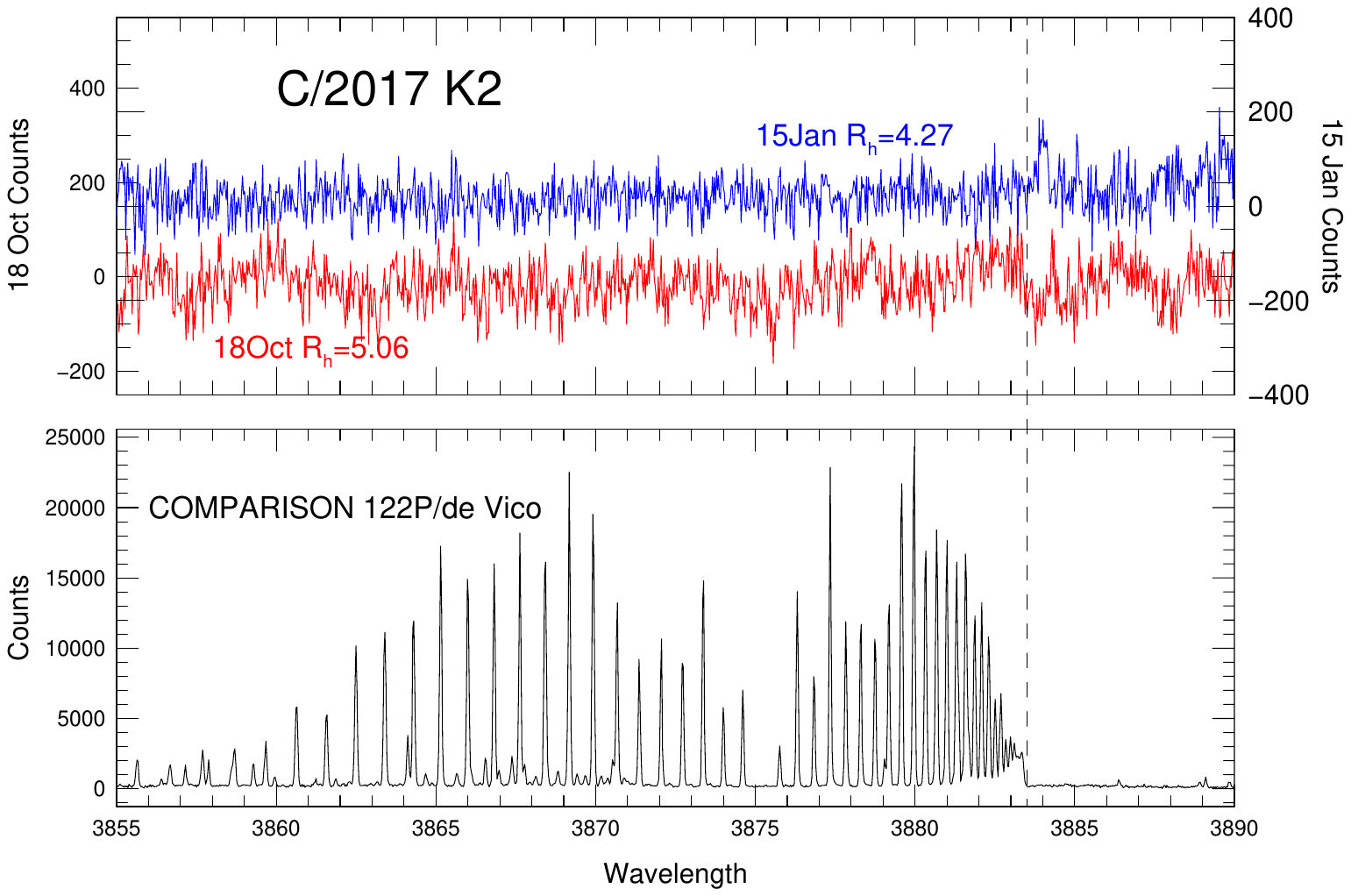}
\caption{The region of the solar-removed CN band is shown for 18 Oct 2021
(scale on left side)
and 15 Jan 2022 (scale on right side) along with the de~Vico spectrum for
comparison.  Note the two K2 scales are offset with respect to one\
another for clarity. The dashed
line denotes the bandhead of the CN band for de~Vico.  There is an apparent emission
just to the red of the bandhead for 15 Jan (with an apparent absorption on
18 Oct).  There are no other features in either K2 spectrum.
\label{jan15CN}}
\end{figure*}

Figure~\ref{jan15CN} shows the solar subtracted spectra of K2 on 18 Oct
2021 and 15 Jan 2022, along with the de~Vico spectrum of the same region. 
Inspection of this figure shows there is an emission in the 15 Jan
spectrum slightly red of where
the bandhead for CN should be. There is an absorption feature at that same
place in the 18 Oct spectrum.  There are no other obvious lines at bluer
wavelengths, nor near where the de~Vico spectrum peaks (the bandhead). 
We interpret this
emission feature in the 15 Jan spectrum as inconvenient noise in the
spectrum (possibly still some residual galaxy spectrum).

\begin{figure}
\centering
\includegraphics[width=0.475\textwidth]{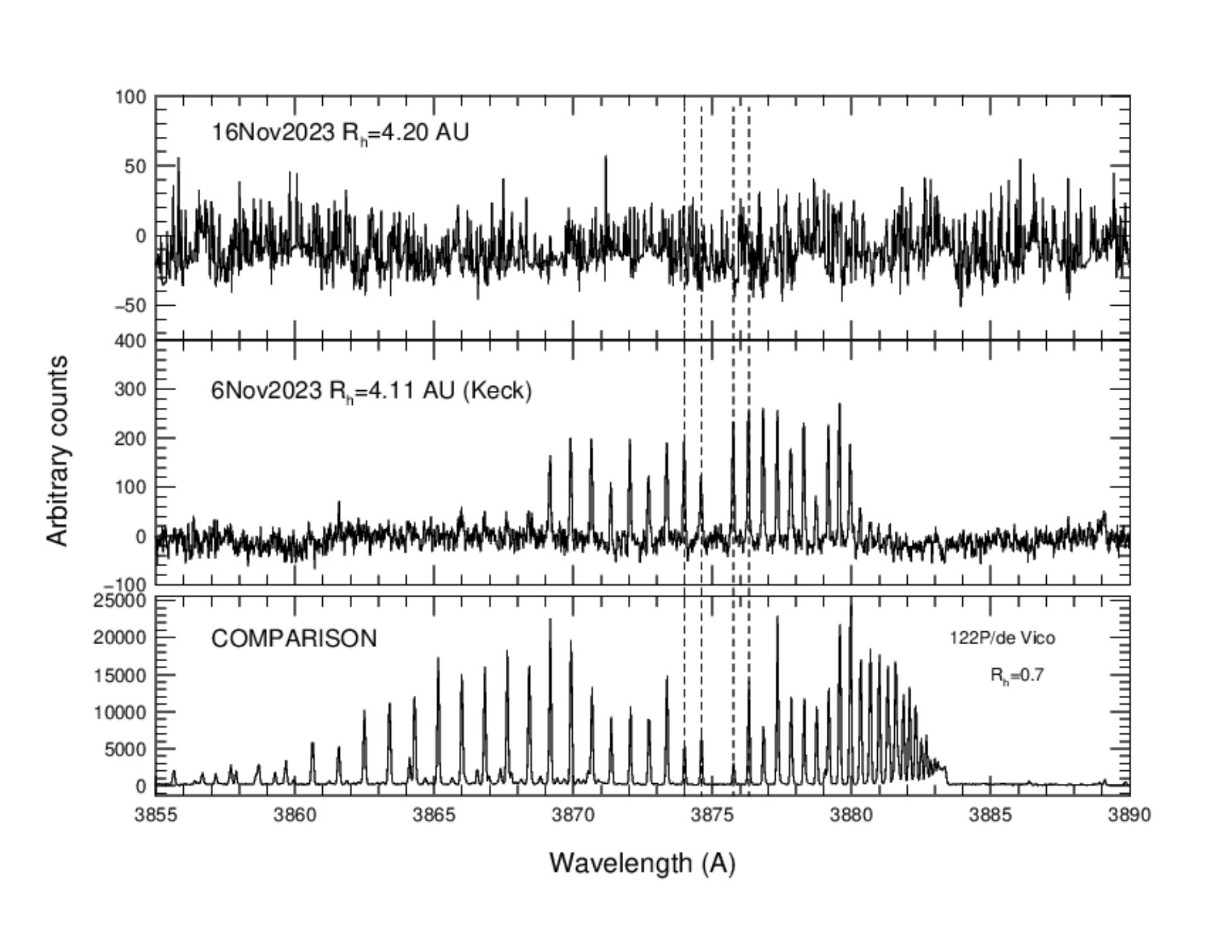}
\caption{The middle panel of this figure shows the detection of CN with Keck at
4.11\,{\sc au} in November 2023.  The top panel shows the lack of detection
of CN at 4.2\,{\sc au} from McDonald ten days later. It is possible that
this lack of
detection is a matter of lack of sensitivity. The bottom panel is the
de~Vico spectrum as a comparison.  
Note that even with Keck we only see the lowest J-level lines for the P and R 
branches and do not see an obvious bandhead.
\label{cnpost}}
\end{figure}

Once the comet moved north again, we were able to obtain more
spectra of the comet, once with Keck HIRESb and once at McDonald.  There was
an obvious CN feature in the 6 Nov 2023 HIRESb data obtained with the comet
at 4.11\,{\sc au} but no other molecules
were detected.  Ten days later, and 0.09\,{\sc au} further from the Sun,
we again observed K2 at McDonald, but did not observe any emission features.
Figure~\ref{cnpost} shows the detection of CN on 6 November 2023 and the
probable non-detection on 16 November 2023.
The Keck telescope has 13.7 times the collecting area of the McDonald 2.7m
telescope and we obtained 2 times the integration time at Keck as we did at
McDonald. Thus, we were much less sensitive on 16 Nov 2023 than on 6 Nov
2023 and we cannot be 10@article{stze2000,title="Theoretical values for the {[O III]} 5007/4959
        line-intensity ratio and homologous cases", author="Storey, P. J.
        and Zeippen, C. J.",journal=mnras,volume=312,pages=813,
        endpage=816,year=2000}

0\% certain that there were no molecular emissions on that date. Table~\ref{results} summarizes what was found for each molecule.

\begin{table}
\caption{Summary of Detected Molecules \label{results}}
\centering
\begin{tabular}{r@{\,}l@{\,}lcccccc}
\hline
\multicolumn{3}{c}{Date} & R$_h$ & CN & C$_{3}$ & CH & C$_{2}$ & NH$_{2}$ \\
\hline
18 & Oct & 2021 & 5.06 & N & N & N & N & N \\
15 & Jan & 2022 & 4.27 & N & -- & -- & -- & -- \\
22 & Apr & 2022 & 3.39 & Y & N? & N & Y? & N \\
09 & May & 2022 & 3.24 & Y & Y & N & Y & N? \\
10 & Jun & 2022 & 2.95 & Y & Y & Y & Y & Y? \\
25 & Jun & 2022 & 2.82 & Y & Y & Y & Y & Y \\
31 & Jul & 2022 & 2.51 & Y & Y & Y & Y & Y \\
06 & Aug & 2022 & 2.47 & Y & Y & Y & Y & Y \\
06 & Nov & 2023 & 4.11 & Y & N & N & N & N \\
16 & Nov & 2023 & 4.20 & N & N & N & N & N \\
\hline
\end{tabular}
\end{table}

\section{Atomic Oxygen Lines in the Spectra}
In addition to the emission spectrum of the molecules, cometary spectra show 
three forbidden atomic oxygen lines in emission.  These are the 5577\,\AA\ 
[O\,$^1$S] line, and the 6300\,\AA\ and 6363\,\AA\ [O\,$^1$D] lines.
\citet{fefe81} showed that these lines do not arise from fluorescence, even
for an extremely active comet.
They concluded that the atomic oxygen lines are produced
by a single
dissociation of H$_{2}$O, CO or CO$_2$ and are prompt emissions (very short
lifetimes).

\citet{decocketal2013} examined the importance of collisional quenching for the strength of the oxygen lines.  They showed that the line ratio is not affected by collisional quenching unless the water production rate was greater than $10^{30}$\,sec$^{-1}$, much higher than the H$_2$O production rate for K2 of around $10^{28}$\,sec$^{-1}$ measured by
\cite{Ejeta_2025}. Similarly, \citet{mckay2016} found no need to include collisional quenching for two
comets (C/2012 K5 (LINEAR) and 290P/J{\"a}ger) with comparable production rates and observing geometry.

The ratio of the
green line flux to the sum of the red line fluxes can yield an
estimation of the importance of H$_{2}$O, CO and CO$_2$ for controlling sublimation.
Since we cannot observe CO$_2$ features from ground-based telescopes due to
CO$_2$ in the Earth's atmosphere, this line ratio can be utilized as a proxy for
measuring the CO$_2$ content of the comet. 
This has been used extensively by many authors (c.f. \citet{mckay12,mckay2013,mckay2016,decocketal2013,decocketal2015}).
The line ratio for dissociation when H$_{2}$O is the parent is around 0.1,
whereas when the source is CO or CO$_2$, the ratio is more likely to be 1 \citep{fefe81} (though \citet{de80b} argued for a path for dissociation of CO$_2$ that could yield an oxygen line ratio of 0.3). 

We automatically observed these three lines in the McDonald Observatory spectra. However, these same lines also appear as emission features in the telluric spectrum. As a result, we can only measure the cometary oxygen line fluxes when they are Doppler shifted sufficiently from the telluric lines, allowing the cometary and telluric features to be distinguished. We had comet/telluric relative Doppler shifts of sufficient magnitude for all of the McDonald spectra obtained between 22 April 2022 and 6 August 2022.

Before we could measure the oxygen lines, we needed to remove other features from the spectra that would affect the measured flux of the oxygen lines. There are strong telluric absorption features near the two red lines, with the region of the 6300\AA\ line containing pairs of O$_2$ lines that are mostly well separated from the cometary oxygen line. The 6363\AA\ region, however, includes many lines that lie close together near the comet line.

We used observations of a rapidly rotating hot star to divide out these features from the comet spectra. Hot stars have few intrinsic spectral lines, and in rapidly rotating stars, these lines are broadened and smeared out to the point of being nearly undetectable; the narrow lines remaining in these spectra are primarily telluric in origin.

\begin{figure}
\centering
\includegraphics[width=0.475\textwidth]{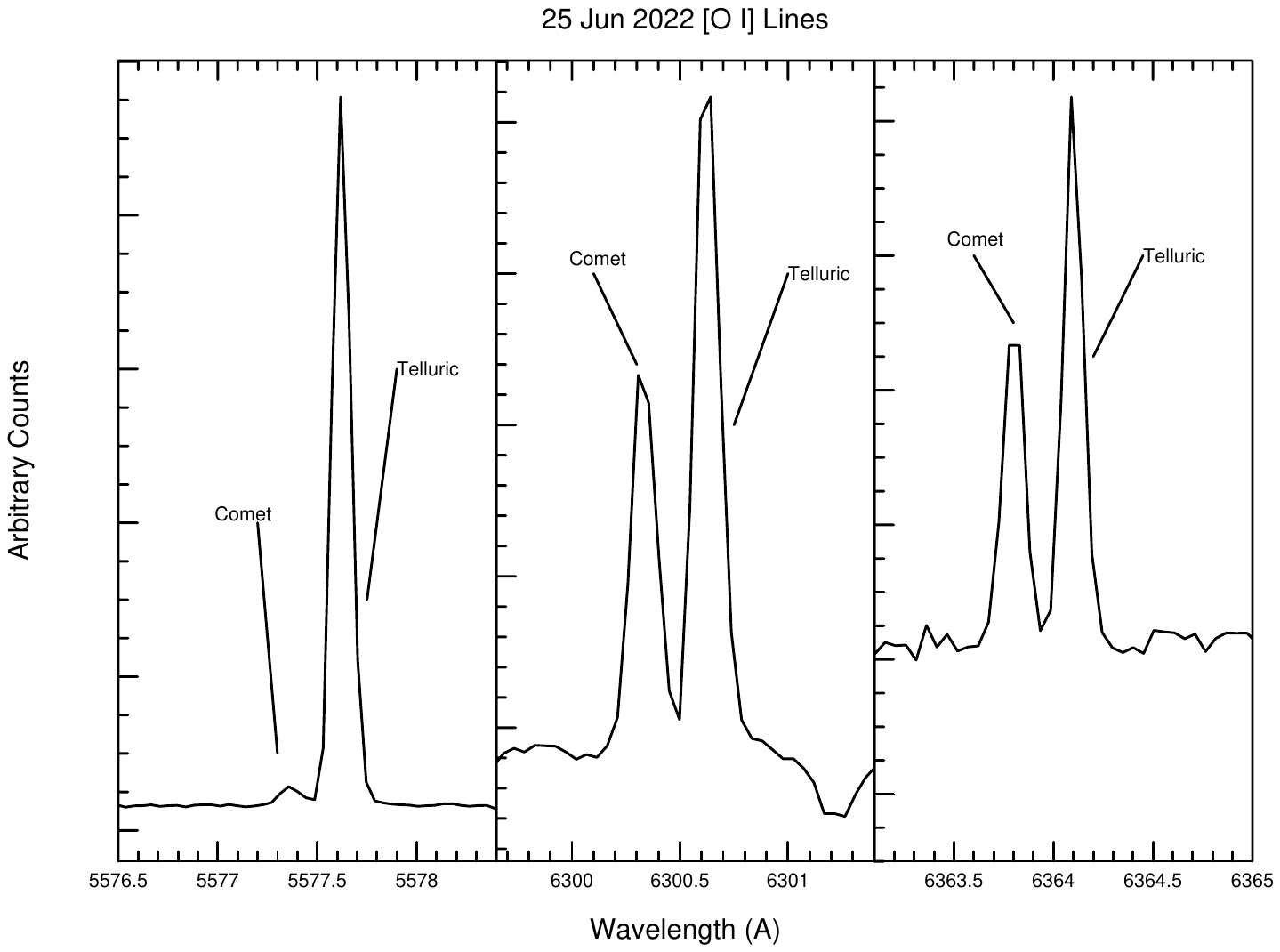}
\caption{The three forbidden oxygen lines are shown for 25 Jun 2022.  Inspection of the three panels shows that the relative strengths 
of the comet lines and the telluric lines are very different for the 5577\AA\ lines versus the 6300 and 6363\AA\ lines.
\label{oxygenplot}}
\end{figure}

Next, we removed the comet’s continuum and absorption spectrum using the solar spectrum, as described above. We then used flux standards to correct for the relative sensitivity across the three spectral orders.
Figure~\ref{oxygenplot} shows examples of the three oxygen lines on 25 June 2022. Both the comet and telluric lines are obvious, but the ratio of the telluric to comet line strengths is very different for the green and red lines. Inspection of this figure shows that our observations of the 5577\AA\ oxygen lines are not affected by the presence
of any possible C$_2$ emission lines that often cause problems with the measurement of the green oxygen line in more active comets. This comet simply has such weak C$_2$ that these lines are absent in our data. 

\citet{decocketal2015} looked at how the line ratio changes with distance to the nucleus.  However, in their study, all of the comets were observed at R$_h \sim1$\,{\sc au} and $\Delta$ much smaller than 1\,{\sc au}. They observed a change in the line ratio for nucleocentric distances $<100$\,km.  The smallest geocentric distance for our observations was 1.86\,{\sc au} on 31Jul 2022. At that distance, our $1.2\times8.2$\, arcsec slit was about $1620\times11060$\,km. We therefore felt it reasonable to ignore collisional quenching.

For most of the usable observation period, the cometary and telluric lines were partially blended. To address this, we deblended them by fitting overlapping Gaussian profiles to the data. Table~\ref{oxygen} shows the values we derived for these line ratios.

\begin{table*}
\centering
\caption{The Oxygen Line Ratios \label{oxygen}}
\begin{tabular}{r@{\,}l@{\,}lcccc}
\hline
\multicolumn{3}{c}{Date} & Heliocentric & Geocentric & Red & Green/Red\_Sum \\
\multicolumn{3}{c}{(UT)} & Distance & Distance & Line & Line \\
 & & & (AU) & (AU) & Ratio & Ratio \\
\hline
22 & Apr & 2022 & 3.39 & 3.00 & 3.15$\pm$0.06 &0.25$\pm$0.007 \\
09 & May & 2022 & 3.24 & 2.64 & 2.92$\pm$0.05 &0.15$\pm$0.005\\
10 & Jun & 2022 & 2.95 & 2.05 & 2.82$\pm$0.02 &0.18$\pm$0.002 \\
25 & Jun & 2022 & 2.82 & 1.89 & 2.93$\pm$0.02 &0.16$\pm$0.002 \\
31 & Jul & 2022 & 2.51 & 1.86 & 3.28$\pm$0.03 &0.13$\pm$0.003 \\
06 & Aug & 2022 & 2.47 & 1.90 & 2.97$\pm$0.01 &0.11$\pm$0.001 \\
\hline
\end{tabular}
\end{table*}

Examination of Table~\ref{oxygen} shows that the line ratio changed a small amount with heliocentric distance, decreasing as the comet approached closer to the Sun.  However, all nights except 22 Apr, when the comet green line was very weak, are approximately the same.
A similar trend of increasing line ratio with increasing heliocentric distance was observed in spectra of comet Garradd 
\citep{mckay2015,decocketal2013},
although over a somewhat smaller range of heliocentric distances. It is possible that the larger ratios at heliocentric distances greater than 3 {\sc au} result from the lower signal-to-noise ratio in those observations, as the comet was fainter when farther from the Sun.

However, all measured values are consistent with the atomic oxygen lines being dissociation products of H$_{2}$O. This finding suggests that even though the comet was active at heliocentric distances beyond 3 {\sc au}, H$_{2}$O may still be the dominant driver of cometary activity, possibly due to H$_2$O crystals in the coma subliming and contributing to the overall line strengths.

\citet{caetal2023} observed the three oxygen lines on two nights (24Jun2022 at 2.83\,{\sc au}
and 2Jul2022 at 2.75\,{\sc au}).  They found values for the oxygen ratio of 0.29 and 0.27 for their two nights and concluded that this showed that the oxygen was a product of CO$_2$ dissociation. Their measured value for the line ratio on 24Jun was almost 2 times higher than our value on 25Jun.  We cannot account for this difference, but we note that they were measuring with a 1\,arcsec aperture while our aperture was $1.2 \times 8.2$\,arcsecs.  Our larger aperture might have diluted the signal a little.  We also note that \citet{caetal2023} derived a ratio of $\sim2.3$ for the two red lines, while physics requires this ratio should be 3 \citep{stze2000}.  Our values for the red line ratios are between 2.8 and 3.3
(see Table~\ref{oxygen} for our values and error bars).  These suggest the amount of measurement error we obtained.
Finally, we note that on all nights, the green line was the same FWHM  (0.12\AA) as the two red lines.  This makes it plausible that the red lines and the green line have the same parent source.

\citet{Hmid_K2} observed K2 with UVES on the VLT on four nights (9 May 2022, 5 Jul 2022, 21 Sep 2022 and 22 Sep 2022). They also observed the three oxygen lines
and were able to measure the ratio of the green line flux to the sum of the red line fluxes.  Their ratios ranged from 0.25 in May when the comet was at a heliocentric distance of 
3.23\,{\sc au} to 0.08 in September when the comet was at a heiocentric distance of 2.11\,{\sc au}. This is  in good
agreement with what we found. 

\section{Discussion}
Our observations began when K2 was bright and extended, but still more than 5\,{\sc au} from the Sun. At this heliocentric distance, we detected no emission features from any gases. This aligns with expectations if H$_{2}$O were the primary driver of ice sublimation and gas release. Indeed, molecular emissions would typically not be expected  to be observed until the comet approached a heliocentric distance of around 3\,{\sc au}, anticipated in late May or early June of 2022.

However, that is \underline{not} what we observed. Instead, we made the first definitive detection of CN gas on 22 Apr 2022, when the comet was at 3.39\,{\sc au}. At that distance, we also observed a possible detection of C$_{2}$ emission.

An explanation for this early gas release is that the nucleus of K2 may not be dominated by H$_{2}$O ice.  
This is a reasonable possibility based on the detection of CO$_2$ by \citet{yangetal2021} at 6.7\,{\sc au}. However, the spectrum of K2 when it was well within a heliocentric distance of 3\,{\sc au} appears quite typical, with no unusual features.
In addition, the oxygen line ratios we observed for K2 are
also compatible with H$_{2}$O being a dominant producer of these oxygen
lines, though \citet{caetal2023} concluded that the oxygen allows for CO$_2$ as a parent.

H$_2$O, CO, and CO$_2$ were detected with JWST when K2 was at 2.35\,{\sc au} \citep{woodwardetal2025}.
The CO and CO$_2$ production was 8--15\% that of the H$_2$O in these observations. They also found that K2 was a hyperactive comet, with a water-ice fraction  $>$86\%.  This would be inconsistent with CO or CO$_2$ controlling most of the activity. It is possible that H$_2$O is involved with controlling the activity of some species, resulting in different species turning on at different heliocentric distances.

Observations over a range of heliocentric distances (3.15 -- 2.35\,{\sc au}) were obtained with the IRTF \citep{Ejeta_2025}. They showed that the ratios of volatiles to water were significantly higher for K2 than for most other Oort Cloud comets. They concluded that K2 was either enriched in these (normally “trace”) volatiles, was depleted in water, or H$_2$O was not fully activated beyond around 2\,{\sc au}.

Our belief that emission line features should turn on at 3\,{\sc au} if the
comet's volatility is controlled by water is based on the temperature at
which H$_{2}$O ice sublimes and the distance from the Sun at which
the comet encounters that temperature.  However, our knowledge of the activity
of comets at this distance is actually quite limited, even with surveys
that have observed a large number of comets, because comets at $>$3{\sc
au} are not that bright.  \citet{ahetal95} reported on photometric
observations of 85 comets. Of those, only 10 were observed at heliocentric distances $>$3{\sc au}, and 4 of those 10 showed no gas emission. Most of the comets observed beyond 3{\sc au} exhibited no detectable gas at heliocentric distances greater than 3.5\,{\sc au}.

\citet{coetal2012} reported on  spectroscopic observations of 130 comets,
with only 15 having observations at $>$3\,{\sc au}.
Of those 15 comets, only 5 had gas detected beyond 3\,{\sc au} and only 2
of those comets (1P/Halley and C/1995\,O1 (Hale-Bopp)) had gas observed at more
than 4\,{\sc au}.  Those two comets were the most productive comets in this
survey.

D. Schleicher and A. Bair (2025, personal communication) have continued collecting photometric observations and expanding the database of \citet{ahetal95}. Through 2021, their database contained observations of 220 comets, with only 24 of those observed at heliocentric distances $>$3.5\,{\sc au}.

These surveys show a bias of {\it not} observing comets outside the distance where we expect H$_{2}$O to sublime.   
Each survey had observations of comets at heliocentric distances $>$3.5\,{\sc au} for only 10\% of their targets.
Therefore, it is difficult to say how unusual K2 really is.  
This is the first passage into the inner Solar System for
this comet.  The well-developed spectrum at larger heliocentric distances may result from more volatile species remaining in the outer layers of the comet, driving some of the activity during its first passage.  A comet making a
subsequent passage into the inner Solar System (which K2 will not do
in our lifetimes) would
have shed this more volatile material and might turn on closer to
the Sun the second time around.  Obviously, this is not something we can observe
because this is essentially K2's only passage near the Sun.

Looking at Table~\ref{results} we can also see that not all species started
to be apparent at the same heliocentric distance. This can be explained in part by the fact that some molecular features are generally easier to detect than others.  It
is not uncommon for CN to be the first molecule whose lines appear above
the continuum, but not generally so far from the Sun. 
If we invoke, again, more volatile species on the surface of the nucleus,
this could indicate that the release of different species is clumpy, with
some pockets of molecules being released at different times.
It is, however, somewhat surprising that we detected CN on the outbound leg at greater than 4\,{\sc au}.

\section{Conclusion}

The activity of comet C/2017 K2 (PanSTARRS) at large heliocentric distances challenges the traditional view that H$_{2}$O ice is the sole driver of early cometary activity and other species near the surface of the nucleus might be driving some of what we observe. Our detection of CN at 3.39\,{\sc au}, with additional possible detections of other species, suggests that more volatile ices such as CO and CO$_2$ may contribute significantly during the comet’s inbound leg. While the coma composition inside 3\,{\sc au} appears typical of water-driven comets, elevated volatile-to-water ratios and JWST data indicate a complex outgassing behavior, possibly related to the comet’s first passage through the inner Solar System. The detection of molecular emissions beyond 4\,{\sc au} on the outbound leg further supports the presence of heterogeneous, clumpy volatile sources on the nucleus and may be indicative that 
a CN parent is something more volatile than water. Taken together, these results underscore the need for more systematic observations of comets beyond 3{\sc au}, and suggest that K2 may represent a broader class of Oort Cloud comets with extended, volatile-driven activity at large distances from the Sun.

\begin{acknowledgements}
This work was performed under NASA Grant 21-SSO21-0007; Grant Number 80NSSC22K0660. This paper includes data obtained at the McDonald Observatory of The University of Texas at Austin. 

Some of the data presented herein were also acquired at the W. M. Keck Observatory, which is a private 501(c)(3) non-profit organization operated as a scientific partnership among the California Institute of Technology, the University of California, and the National Aeronautics and Space Administration. The Keck Observatory was made possible by the generous financial support of the W. M. Keck Foundation.
The authors wish to recognize and acknowledge the very significant cultural role and reverence that the summit of Maunakea holds within the Native Hawaiian community. We are truly privileged to have the opportunity to conduct observations from this mountain.
\end{acknowledgements}

\facility{McDonald Observatory (2.7m Tull Coude Spectrograph)}
\facility{Keck:I (HIRES)}

\bibliographystyle{aasjournal}

\end{document}